\documentclass[hidelinks,conference]{IEEEtran}
\IEEEoverridecommandlockouts
\usepackage{cite}
\usepackage{amsmath,amssymb,amsfonts}
\usepackage{algorithmic}
\usepackage{graphicx}
\usepackage{textcomp}
\usepackage{xcolor}
\usepackage{csquotes}
\usepackage{hyperref,filecontents}
\usepackage[inline]{enumitem}

\usepackage{cleveref}

\usepackage[acronym]{glossaries}
\usepackage{glossaries-extra}
\setabbreviationstyle[acronym]{long-short}

\usepackage{mathastext}

\global\marginparsep=0pt
\global\marginparwidth=47pt
\global\marginparpush=10pt

\newacronym{aas}{AAS}{Asset Administration Shell}
\newacronym{ai}{AI}{Artificial Intelligence}
\newacronym{api}{API}{Application Programming Interface}
\newacronym{ar}{AR}{Augmented Reality}
\newacronym{bpmn}{BPMN}{Business Process Model and Notation}
\newacronym{cpps}{CPPS}{Cyber-Physical Production Systems}
\newacronym{cv}{CV}{Computer Vision}
\newacronym{dt}{DT}{Digital Twin}
\newacronym{gui}{GUI}{Graphical User Interface}
\newacronym{hmi}{HMI}{Human Machine Interface}
\newacronym{http}{HTTP}{Hypertext Transfer Protocol}
\newacronym{ibpt}{IBPT}{Industrial Business Process Twin}
\newacronym{iira}{IIRA}{Industrial Internet Reference Architecture}
\newacronym{iot}{IoT}{Internet of Things}
\newacronym{it}{IT}{Information Technology}
\newacronym{mes}{MES}{Manufacturing Execution System}
\newacronym{nist}{NIST}{National Institute of Standards \& Technology}
\newacronym{nlp}{NLP}{Natural Language Processing}
\newacronym{opcua}{OPC~UA}{Open Platform Communications Unified Architecture}
\newacronym{ot}{OT}{Operational Technology}
\newacronym{plc}{PLC}{Programmable Logic Controller}
\newacronym{rami40}{RAMI4.0}{Reference Architecture Model Industry~4.0}
\newacronym{rpc}{RPC}{Remote Procedure Call}
\newacronym{rl}{RL}{Reinforcement Learning}
\newacronym{scada}{SCADA}{Supervisory Control and Data Acquisition}
\newacronym{soa}{SOA}{Service Oriented Architecture}
\newacronym{sos}{SoS}{System of Systems}
\newacronym{tcpip}{TCP/IP}{Transmission Control Protocol/Internet Protocol}
\newacronym{uml}{UML}{Unified Modeling Language}
\newacronym{zvei}{ZVEI}{German Electrical and Electronic Manufacturers' Association}

\begin{document}

\title{Digital Twins of Business Processes as Enablers for IT / OT Integration\\
\thanks{Supported by the Christian Doppler Research Association (JRC ISIA). This preprint has not undergone peer review or any post-submission improvements
or corrections.}
}

\author{\IEEEauthorblockN{Hannes Waclawek, Georg Schäfer, Christoph Binder, Eduard Hirsch, Stefan Huber}
\IEEEauthorblockA{Salzburg University of Applied Sciences, 5412 Puch, Austria\\
\{firstname.lastname\}@fh-salzburg.ac.at}}

\maketitle

\begin{abstract}

    The vision of Industry~4.0 introduces new requirements to \gls{ot} systems.
    Solutions for these requirements already exist in the \gls{it} world,
    however, due to the different characteristics of both worlds, these
    solutions often cannot be directly used in the world of \gls{ot}.
    We therefore propose an \gls{ibpt}, allowing to apply methods of one world
    to another not directly but, instead, to a representation, that is in
    bidirectional exchange with the other world.
    The proposed \gls{ibpt} entity acts as an intermediary, decoupling the
    worlds of \gls{it} and \gls{ot}, thus allowing for an integration of
    \gls{it} and \gls{ot} components of different manufacturers and platforms.
    Using this approach, we demonstrate the four essential Industry~4.0 design
    principles information transparency, technical assistance, interconnection
    and decentralized decisions based on the gamified Industry~4.0 scenario of
    playing the game of Nine Men's Morris.
    This scenario serves well for agent based \gls{ai}-research and education.
    We develop an \gls{opcua} information and communication model and
    then evaluate the \gls{ibpt} component with respect to the different views
    of the \gls{rami40}.

\end{abstract}

\begin{IEEEkeywords}
    IT/OT Integration, Digital Twin, RAMI4.0, OPC~UA, SOA, Information Modeling, Industrial Business Process Twin
\end{IEEEkeywords}

\section{Introduction} \label{sec:introduction}
\subsection{Motivation} \label{subsec:motivation}

Systems of \gls{it} and \gls{ot} systems possess inherently different
characteristics.
In~\cite{stouffer2015}, the \gls{nist} lists geographic distribution, safety
and availability, proprietary communication protocols and software platforms,
among others, as typical aspects that need to be considered when designing
\gls{ot} systems in contrast to \gls{it} systems.
Furthermore, the vision of Industry~4.0 introduces new requirements to the
world of \gls{ot}.

As outlined by Hermann in~\cite{Hermann2016}, interconnection, information
transparency, decentralized decisions and technical assistance are the four key
design principles of Industry~4.0 systems.
Solutions for these requirements already exist in the \gls{it} world, however,
due to the different characteristics of both worlds, such
solutions often cannot be directly used in the world of \gls{ot}.\@
Interconnection in Industry~4.0 requires the use of common communication
standards, which we find in the form of the IP protocol suite (IP, TCP, UDP)
and web protocols, like \gls{http} and WebSocket, in the world of \gls{it}.
Contrary to this, as documented in~\cite{stouffer2015}, industrial control systems
currently rely on the use of many proprietary communication protocols.

Information transparency allows for data to be analyzed, which, in the field of
data science, is performed through \gls{ai} or statistical methods that require
\gls{it} software. A holistic semantic data modeling approach is beneficial for
applying these methods. However, proprietary data formats are prevalent in
\gls{ot} systems running hardware of different platforms and manufacturers.
Also, running \gls{ai} methods for data analysis using state-of-the-art \gls{ai}
frameworks like TensorFlow often requires computing power that is not available
on \gls{ot} workstations running time-critical control tasks.

In order to allow decentralized decisions, tasks and information are delegated
from the machine to a higher level, where decentralized decision makers (humans
or programs) can use the information to solve problems or optimize processes. On
the one hand, today's \gls{scada} and \glspl{mes} already realize this
functionality on a factory level. On the other hand, a holistic, multi-factory
approach considering geographic distribution requires the use of common
communication standards allowing information exchange of different systems over
the Internet. As mentioned earlier, this contradicts the many proprietary data
formats which are prevalent in \gls{ot} systems.

Also, \gls{ai}-based decision-making (e.g.\ via \gls{rl}) can further enhance
and automate decisions. This implies \gls{ai} and statistical methods that
require \gls{it} software. As outlined before, however, it is often not feasible
running state-of-the-art \gls{ai} frameworks like TensorFlow on \gls{ot}
workstations. Technical assistance benefits from implementing modern HMIs
allowing human-robot interaction in an intuitive way that is beneficial for the
use in industrial environments.
\gls{nlp} solutions addressing this problem, like Amazon's Alexa, or \gls{ar}
devices, like Microsoft's Hololens, offer benefits over classical input devices
like keyboards, mice or touchscreens in regard to hygienic, safety (e.g.\ visual
attention) or haptic (e.g.\ wearing of gloves) concerns found in production
facilities. However, these solutions also require \gls{it} software and often
possess limited capabilities without an active Internet connection, since voice
or image recognition tasks often run in the cloud.
Due to security concerns, however, Internet access often is not available from
the \enquote{shop floor} level of a production facility.

This line of argumentation shows that a lot of solutions to the key challenges
of Industry~4.0 systems already exist in the world of \gls{it}, but, generally
speaking, these \gls{it} solutions cannot be directly applied to \gls{ot}
systems.

\subsection{Contributions}
\label{sec:contributions}

We therefore propose an \gls{ibpt} entity, allowing to apply methods of one
world to another not directly but, instead, to a representation, that is in
bidirectional exchange with the other world. 
In this way, it is interconnecting the worlds of \gls{it} and \gls{ot} in a
novel way. By describing the \gls{ibpt}, in this work, our contributions are:
\begin{itemize}
    \item Illustration of \glspl{dt} of industrial business processes for
    interconnecting \gls{it} and \gls{ot} systems.
    \item Novel way of reducing \gls{ot} system complexity from an \gls{it}
    perspective by providing unified interfaces at a business process
    abstraction level.
    \item \gls{soa}-centric business case modeling approach for industrial
    settings, that allows for an easy integration of \gls{it} and \gls{ot}
    components.
\end{itemize}

Game-based scenarios are very suitable for testbeds utilized for
\gls{ai}-related research at universities.
For one, literature suggests that gamification is beneficial for
education~\cite{nah2014gamification}.
Also, game-based scenarios serve well for agent-based \gls{ai} research, since
environments, states and actions are well-defined.
In an attempt to create a versatile testbed for \gls{ai}-related research that
can be utilized in education at the Salzburg University of Applied Sciences,
we therefore map the principles of order-driven production to the game based
scenario of playing Nine Men's Morris in section~\ref{sec:scenario}.
We provide an argumentation for the development of the \gls{ibpt} 
in section~\ref{sec:digital-twin-business-process} and develop an
\gls{opcua} information model for the presented scenario in
section~\ref{sec:information-modeling-and-software-design}.
We then evaluate the \gls{ibpt} component with respect to the different views
of the \gls{rami40} in section~\ref{sec:our-solution-in-context-rami4.0}.

\subsection{Related work}
\label{sec:related-work}
Current work in the subjects of digitalization and application of intelligent
services in industrial scenarios strives for the transition of the ISA-95 model,
which is the de-facto standard in the industry today, to approaches satisfying
the needs of Industry~4.0 settings developed in the last decade.
This led to models like \gls{rami40}~\cite{Zvei2015},
\gls{iira}~\cite{Lin2015}, or C5~\cite{Lee2015}.
Modern frameworks try to implement the principles of these models using
\gls{soa}-based approaches.
One of the biggest research projects in the EU in this field led to the
development of the Arrowhead Framework described in~\cite{Delsing2017}, which
uses a \gls{soa}-based \gls{sos} approach to satisfy the needs
of Industry~4.0 settings.
However, this framework targets automation systems in general, from classical
production plants over Smart Grids to Smart Cities, while, in this work, we want
to focus on the Industry~4.0 production setting.
There is a lot of work on \glspl{dt} in production scenarios twinning physical
entities, like production units, however, literature is sparse on \glspl{dt}
twinning industrial business processes.
This is reflected in survey papers like~\cite{Kritzinger2018}
and~\cite{jones2020}.

\section{\gls{ibpt}: Industrial business process twin for decoupling of IT and OT concerns}
\label{sec:digital-twin-business-process}

In his essay~\cite{dijkstra1982}, Dijkstra describes the well-known principle
of \enquote{separation of concerns} as a technique for ordering thoughts
when assessing a problem. 
One attempt to separate the concerns of \gls{it} and \gls{ot} in this way is via
the separation of \enquote{office floor} and \enquote{shop floor} levels in
production businesses described in~\cite{Heidel2019}, since \gls{ot} equipment
is only found on the latter.
Looking at the levels of the \gls{rami40} \emph{Hierarchical} perspective
depicted in Fig.~\ref{fig:hierarchical}, which shows the vertical integration within
a production facility, we can therefore attribute its lower five levels to
\gls{ot}, since they comprise the \enquote{shop floor} of the production
facility.
If we now want to link the worlds of \gls{it} and \gls{ot} via a representation,
it is thus a natural approach to place it at the intersection between the
\emph{Enterprise} and \emph{Workstation} layers. 

In order to abstract functionality of the \gls{ot} subsystem in the \emph{Hierarchical}
perspective, we follow the \emph{Facade} design pattern described by Gamma et
al. in \cite{gamma1995} and provide unified interfaces for interacting with the
\gls{ot} system. This allows for an easier usage of the \gls{ot} system from an
\gls{it} perspective. We do so by implementing abstraction of functionality
throughout the layers of the \gls{rami40} \emph{Architectural} perspective shown
in Fig.~\ref{fig:architectural}. This leaves us with centering our modeling
approach around business processes at the highest level, since business
processes reside on the \emph{Enterprise} segment of the \emph{Business} layer.
A core idea of the \gls{soa} paradigm is the centering around business
activities~\cite{openGroup2009}.
Considering a software implementation for the abstract representation we want
to realize, among available architectural styles, \gls{soa} therefore is a
highly suitable candidate.


\begin{figure}[tb]
    \centerline{\includegraphics[width=\linewidth]{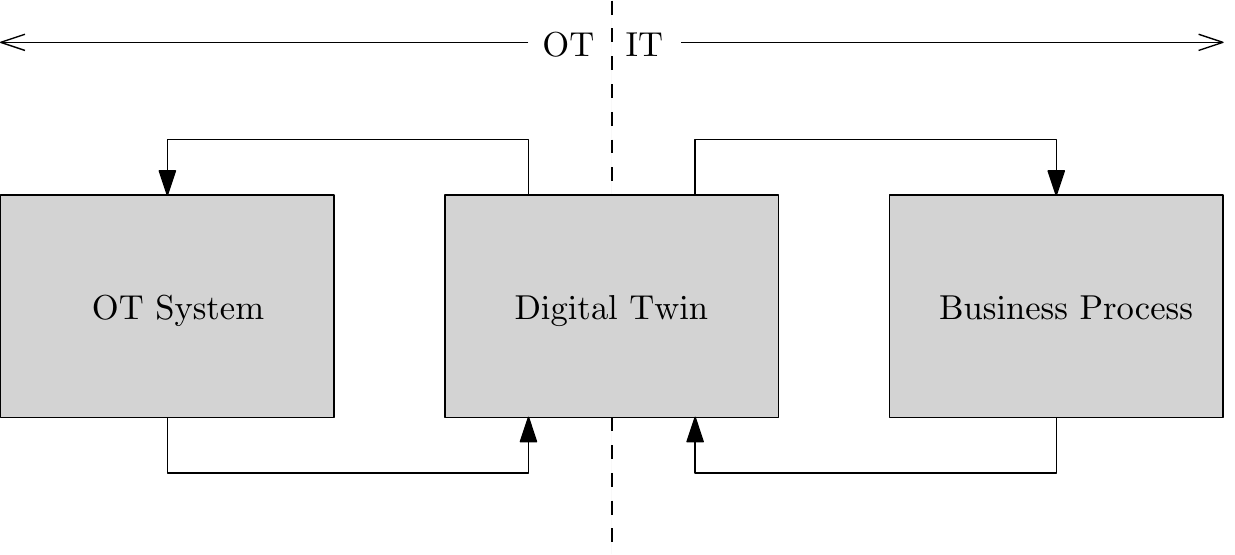}} \caption{The
    \gls{ibpt} concept enables an automated, bidirectional exchange between
    \gls{ot} systems and industrial business processes. It furthermore decouples
    \gls{ot} and \gls{it} concerns via its \gls{dt} entity acting as an
    intermediary/orchestrator between both worlds.}
    \label{fig:ibpt}
\end{figure}

In~\cite{Kritzinger2018}, Kritzinger et al.\ categorize digital representations
according to the type of data flow between the object to mirror and the
mirroring entity:
\begin{itemize}
  \item If data exchange occurs only manually, the mirror is called a digital
    model.

  \item If an automatic data exchange to the mirror is performed, the mirror is
    called digital shadow.

  \item If an automatic data exchange is done in both directions, the mirror is
    called a \gls{dt}.
\end{itemize}

For an automated, bidirectional exchange between \gls{it} and \gls{ot}, our
representation therefore needs to form a \gls{dt}. This, in turn, requires a
suitable communication protocol to establish a bidirectional communication
channel. Also, considering the form of data sent across that channel, we require
a model of information that can be interpreted and understood by both sides in
an attempt to establish information transparency. In~\cite{Heidel2019},
\gls{opcua} is described as the preferred Industry~4.0 communication standard
for the operating phase of production assets. Along with a TCP/IP-based
communication protocol, \gls{opcua} also offers extensive information modeling
capabilities. The OPC~Foundation, the consortium standardizing and advancing
\gls{opcua} technology, offers so-called \emph{Companion Specifications} for
different application domains, like CNC systems or robotics
in~\cite{opcfoundation2022}. In addition, vendors can create their own
extensions as standalone models or on top of these base companion information
models. Considering this, we can utilize \gls{opcua} as a middleware for both,
communication and information modeling, for the purpose of realizing a \gls{dt}
of a business process. As outlined by Mahnke in~\cite{Mahnke2009}, \gls{opcua}
enables a \gls{soa} design approach. As mentioned earlier, this kind of software
architecture is highly suitable for our basic idea of abstracting interfaces of
our \gls{dt} entity to the level of business processes, as it is centered on
modeling business activities. Also, \gls{soa} is recommended as a mechanism to
foster integration capabilities of Industry~4.0 systems~\cite{Heidel2019}.

Following this line of argumentation, the \gls{ibpt} component we propose in
this work represents the \gls{dt} of an industrial business process and is
placed at the intersection between the \emph{Enterprise} and \emph{Workstation}
layers of the \gls{rami40} \emph{Hierarchical} perspective. It is in
bidirectional exchange with both the \gls{ot} and \gls{it} world and acts as an
intermediary/orchestrator between the worlds of \gls{it} and \gls{ot}, thus
addressing challenges for the implementation of \gls{it} methods in \gls{ot}
systems. This principle is shown in Fig.~\ref{fig:ibpt}. \gls{ot} system
complexity is reduced from an \gls{it} perspective by providing unified
interfaces at a business process abstraction level. The \gls{ibpt} entity uses
\gls{opcua} as holistic communication and information modeling platform and
follows the \gls{soa} design approach. 

\section{Developing an \gls{ibpt} Industry~4.0 scenario}
\label{sec:scenario}

In order to be able to describe the \gls{ibpt} component in a specific context,
in this section, we develop a test bed for an Industry~4.0 scenario.
A scenario, in this context, is a specific set of use cases, actors and workflows 
in which our \gls{ibpt} entity is used. 
As a basis, there are several Industry~4.0 scenarios listed in~\cite{Heidel2019}.
The scenario \enquote{flexible factory} contains some elements that are suitable
for our approach.
The scenario describes a combined \gls{mes} / system
integrator unit called \enquote{Production Manager}, which takes production
orders from customers and orchestrates a set of connected production units to
assemble articles.
Like in this scenario, our proposed \gls{ibpt} entity is acting
as an intermediary between the business and production level and an
orchestrator of the latter.

\begin{figure}[tb]
    \centerline{\includegraphics[width=\linewidth, page=2]{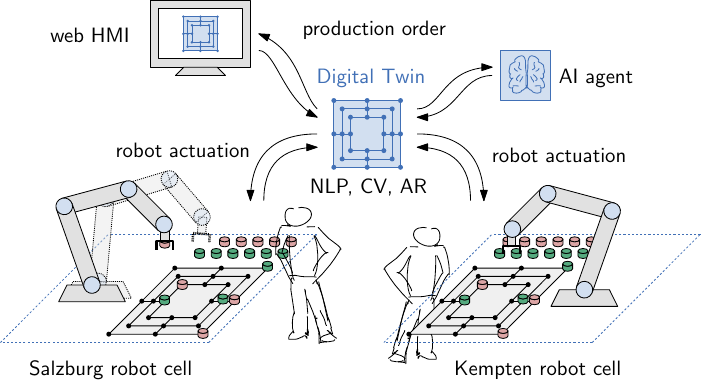}}
    \caption{Schematic representation of the Nine Men's Morris \gls{ibpt} Industry~4.0 scenario.}
    \label{fig:boardgame-concept}
\end{figure}

Game-based scenarios are very suitable for testbeds utilized for
\gls{ai}-related research at universities.
For one, literature suggests that gamification is beneficial for
education~\cite{nah2014gamification}.
Also, game-based scenarios serve well for agent-based \gls{ai} research, since
environments, states and actions are well-defined.
In an attempt to create a versatile testbed for \gls{ai}-related
research that can be utilized in education at the Salzburg University of
Applied Sciences, in this work, we therefore want to map the principles of
order-driven production to the game based scenario of playing Nine Men's
Morris.
This game was chosen due to its simplicity and tokens that can be easily picked
up by different kinds of grippers and distinguished via \gls{cv} systems.
This scenario serves as a test bed for different \gls{ai}-related research
objectives at the Salzburg University of Applied Sciences.
In cooperation with the Kempten University of Applied Sciences, this testbed
was extended to enable geographically distributed communication.
Fig.~\ref{fig:boardgame-concept} schematically shows the actors contributing to
this scenario.

Our proposed setup consists of two physical game boards with corresponding game
tokens located in geographically distributed robot cells.
Robots of different types -- an articulated, USB-controlled robot arm in Salzburg
versus a \gls{plc}-controlled delta robot in Kempten -- interact with the game boards
and demonstrate the integration of machines of different manufacturers and
platforms.
They establish a designated game state, which is represented via token
positions on the respective game boards.
The game state can be changed by performing game moves.
We can identify the following aspects of our proposed \gls{ibpt} entity:

\begin{enumerate}[label=(\Roman*)]
    \item Our \gls{ibpt} represents the business process of a
    production order, which, in our example, is performing a game move.
    \item Humans or \gls{ai} agents can place production orders for changing the game
    state. The \gls{ibpt} entity orchestrates connected production units,
    which, in this scenario, are the robot cells playing the game of Nine Men's
    Morris.
    \item The \gls{ibpt} entity abstracts both elements from the business
    and factory level, while bidirectionally mirroring the state of its production
    units and aggregating information into a holistic representation of the
    abstract business process \enquote{Nine Men's Morris game move}.
\end{enumerate}

We want to demonstrate these aspects in an Industry~4.0 context.
In~\cite{Hermann2016}, Hermann et al.\@ identify
\begin{enumerate*}[label=(\roman*)]
    \item information transparency, \label{enum:designprinciples_transparency}
    \item technical assistance,  \label{enum:designprinciples_techassistance}
    \item interconnection and \label{enum:designprinciples_interconnection}
    \item decentralized decisions\label{enum:designprinciples_decentralization}
\end{enumerate*} as the four Industry~4.0 design principles.
We demonstrate~\ref{enum:designprinciples_transparency} by implementing a
holistic \gls{opcua} data model enabling common data semantics. 
Principle~\ref{enum:designprinciples_techassistance} is shown in a virtual and
physical way by incorporating modern \glspl{hmi} like \gls{nlp} or \gls{ar} for
human-machine interaction.
This requires the integration of \gls{it} (e.g.\@ Google Speech API,
Microsoft Hololens) and \gls{ot} components (e.g.\@ \glspl{plc}) of different
manufacturers, platforms and unknown common software architecture.
This situation allows us to show that our proposed \gls{ibpt}
entity acting as an intermediary, decoupling the worlds of
\gls{it} and \gls{ot}, allows for an integration of \gls{it} and \gls{ot} components
without having to alter existing software packages.
Principle~\ref{enum:designprinciples_interconnection} implies communication
between factories, which brings the challenge of distributed manufacturing
across geographical borders and communication via standard protocols over the
Internet.
This is also true for aspect~\ref{enum:designprinciples_decentralization}, which
implies controlling parts of production processes remotely. Both aspects are
demonstrated by interconnecting the distributed robot cells in our scenario via
the TCP/IP-based \gls{opcua} protocol, which enables communication over the
Internet despite possible proprietary protocols which are utilized on a lower,
factory device level. 

\begin{figure}[tb]
    \centerline{\includegraphics[width=1.0\linewidth, page=11]{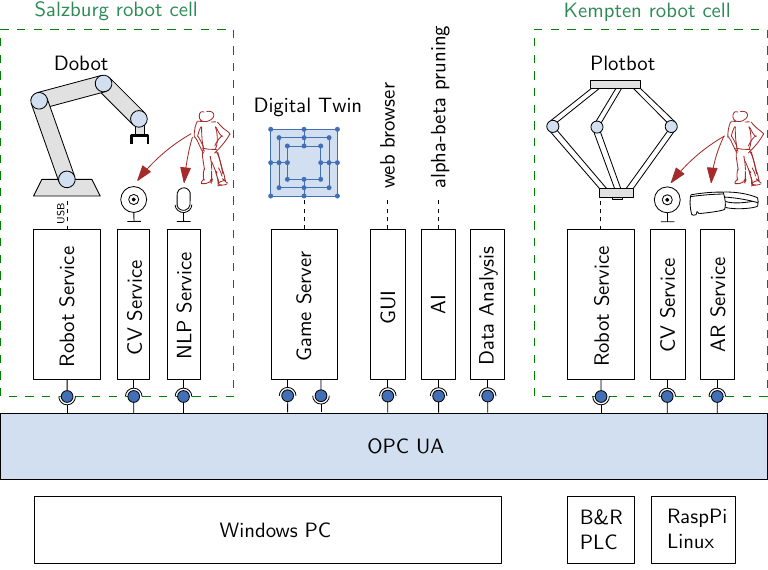}}
    \caption{\gls{soa} for the Nine Men's Morris \gls{ibpt} scenario consisting
    of components of different manufacturers and platforms.}
    \label{fig:boardgame-SOA-architecture}
\end{figure}

Humans or \gls{ai} agents can place production orders for changing the game
state, which demonstrates aspect~\ref{enum:designprinciples_decentralization}.
Fig.~\ref{fig:boardgame-SOA-architecture} shows how this is incorporated into a
\gls{soa} with \gls{opcua} as a middleware for information modeling and
communication at its core. The \emph{Game Server} entity represents the concrete
\gls{ibpt} instance for this \gls{soa}.

\section{Information modelling and software design}
\label{sec:information-modeling-and-software-design}

Since the main goal of our proposed \gls{ibpt} is to enable \gls{it} /
\gls{ot} integration, our software design has to allow for the utilization of
\gls{it} and \gls{ot} solutions of different manufacturers and platforms.
This includes, for instance, different kind of robots (e.g.\@ serial vs.
parallel kinematics, physical vs.\@ virtual robots,\dots), as well as
\glspl{plc}, \glspl{hmi} or \gls{ai} technology of different manufacturers
running on different platforms.

As outlined in~\cite{tanenbaum2017}, one way to enable services in distributed
systems to act independent of their environment is by designing interfaces
that conceal implementation details.
This so-called \emph{encapsulation} makes individual services more
self-contained and is a core principle of \glspl{soa}.
Via encapsulation, device specific implementation details can be solved at a
device level, while interfaces remain valid for all consumers within the
software architecture.
This principle will be one of the core elements of our proposed design,
as we abstract the communication/information model to game logic.
This means that, as an example, instead of sending target coordinate positions
or poses to robots, we solely send abstract game moves.
In this way, the physical properties of the game board (e.g.\@ game field
coordinates) can be defined at a robot cell level, while still meeting the
requirements of our information model.


\subsection{OPC~UA Data Model}
\label{subsec:data-model}

As shown in Fig.~\ref{fig:boardgame-SOA-architecture}
and~\ref{fig:components-diagram}, components are connected using uniform
\gls{opcua} interfaces.
A core idea of the \gls{soa} paradigm is the centering around business
activities~\cite{openGroup2009}.
This business-centric nature of \glspl{soa} calls for an information model
that is centered around the business asset in question, which, in our case, is
the process \enquote{Nine Men's Morris game move}.
We therefore deduct communication to the level of playing the game of Nine
Men's Morris.
This means that, e.g., for triggering token placement at a robot cell, connected
production units receive \texttt{GameMove}s including a
\texttt{FromField} and a \texttt{ToField}, defining a move (e.g.\@ from
\enquote{A1} to \enquote{A7}).
The process of translating this command into robot specific target coordinates
and movement is left to the robot service in question, with specific
geometries and robot capabilities.
This approach increases interoperability, as sending robot specific commands
would require the adaption of the communication model for each robot type.
This abstract data model is realized using \gls{opcua}, which allows us to
increase encapsulation of services by designing interfaces that conceal
implementation details and use a common TCP/IP-based communication protocol.

The foundation of the game Nine Men's Morris is the board with its tokens.
The board is a two-dimensional grid with a total of 24 game fields as well as 18
fields for storing game tokens.
We define an \gls{opcua} data model using enumerations.
The \texttt{GameField} enumeration includes the fields A1 to G7 as well as Tray1
and Tray2.
Tray management is left to the robot cell in question, which means that a game
move from or to a tray does not specify concrete tray positions.
Another enumeration is \texttt{GameFieldOccupationState}, which determines
whether player one, player two or no player token is located on a
\texttt{GameField}.
The enumeration \texttt{PlayerRole} defines player one, player two and
observers.
The latter can be used to not actively take part in the game, but to receive
updates on the current game state, to e.g.\@ display it in a \gls{gui}.

Using these base enumerations, we define several \gls{opcua} data structures.
A \texttt{GameMove} comprises a source and destination \texttt{GameField}.
A \texttt{GameFieldState} comprises a \texttt{GameField} along with its
\texttt{GameFieldOccupationState}.
The \texttt{GameBoard} is an array of 24 \texttt{GameFieldStates}.
The \texttt{GameState} consists of the \texttt{GameBoard} as well as a
\texttt{PlayerRole} to indicate which player is next.
This completes our \gls{opcua} data model types.

\subsection{Architecture}
\label{subsec:architecture}

\begin{figure}[tb]
    \centerline{\includegraphics[width=1.0\linewidth]{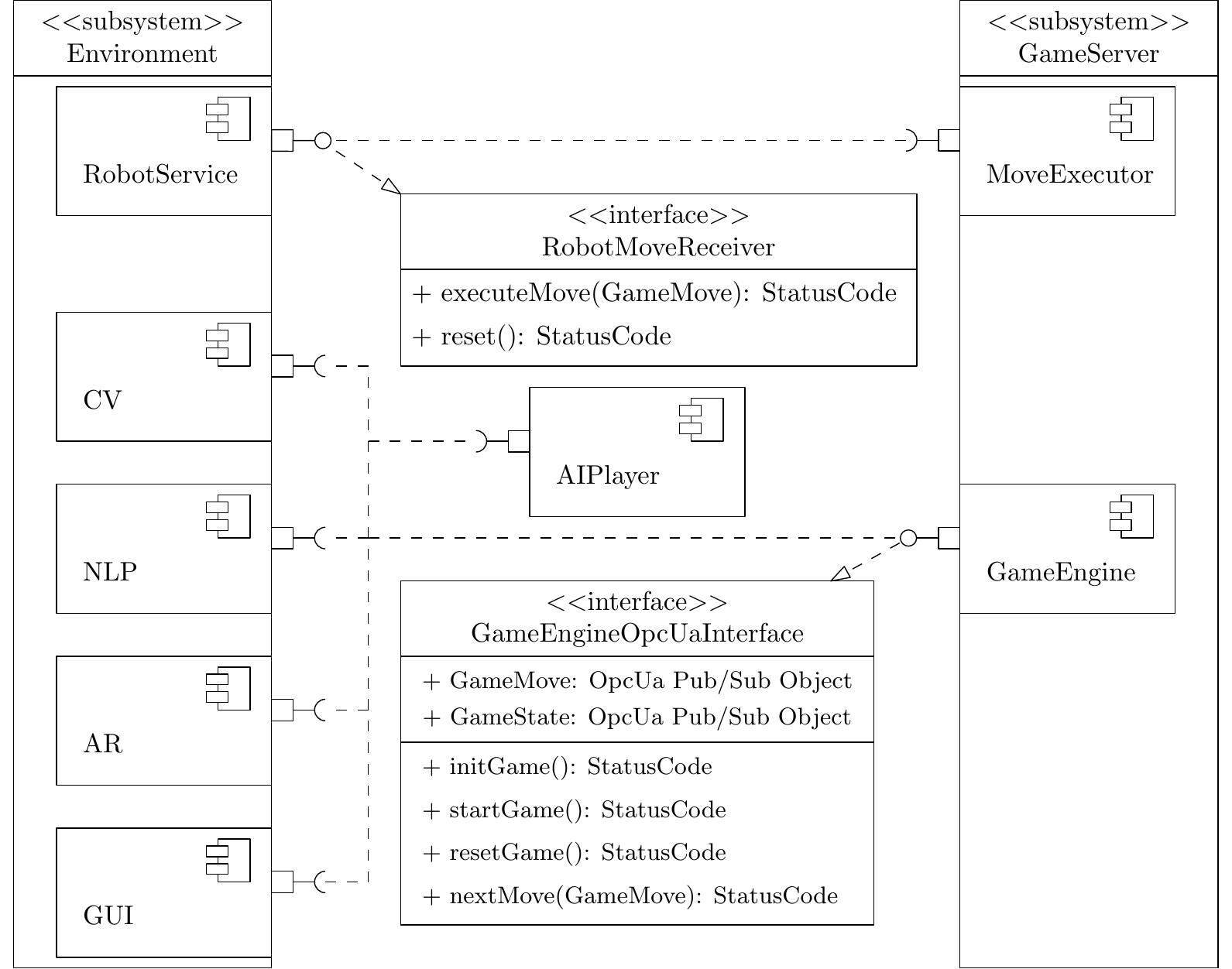}}
    \caption{Component diagram of \gls{soa} for the Nine Men's Morris \gls{ibpt} concept.}
    \label{fig:components-diagram}
\end{figure}

In order to realize a \gls{soa} for our \gls{ibpt} concept, we define several
software components.
Fig.~\ref{fig:components-diagram} shows the component diagram of the resulting
software design.

The core device of the architecture is the \texttt{GameServer}, which handles
communication with different \texttt{Environment}s and \gls{it} components 
and represents the concrete \gls{ibpt} instance.
The \texttt{MoveExecutor} component handles connections to production units.
An environment consists of robot cells with physical or virtual robots, and
several \gls{hmi} services like \gls{cv}, \gls{nlp} or \gls{ar}.
Further components include \texttt{AIPlayer}s and interfaces for human
players, like \glspl{gui}, voice and \gls{ar} interfaces or \gls{cv}
systems.
Communication between the components is realized using \gls{opcua}.
In order to comply with \gls{opcua} standard service calls defined
in~\cite{iec62541-4}, which is the norm describing all interactions between
\gls{opcua} clients and servers via \gls{opcua} services, all \glspl{rpc} defined 
in our system return status codes defined in this norm.

The \texttt{RobotMoveReceiver} interface provides the function
\texttt{executeMove}, which takes a \texttt{GameMove} as an argument in order to
perform token placement.
The process of translating the \texttt{GameMove} into robot specific target coordinates
and movement is left to the robot service in question.
The function \texttt{reset} allows to establish an initial state via function call.
The \texttt{GameEngine} component offers \gls{opcua} Pub/Sub nodes
\texttt{GameMove} and \texttt{GameState} to provide the current state of the
game field as well as the next valid game move to execute on production units.
Function \texttt{nextMove} can be called with a \texttt{GameMove} as an argument
to indicate which move a player wants to take next.
If the move is valid in the current game state, the Pub/Sub nodes
\texttt{GameMove} and \texttt{GameState} are updated to indicate a valid move
along with the new \texttt{GameState} to registered clients.
Functions \texttt{initGame}, \texttt{startGame} and \texttt{resetGame} are used
for game administration.

\section{A RAMI4.0 perspective}
\label{sec:our-solution-in-context-rami4.0}

In an effort of finding a suitable model for describing value chains in
Industry~4.0, the \gls{zvei} introduced the reference architecture model
\gls{rami40}~\cite{Zvei2015}.
In the following chapters, we will use this three-dimensional layer model to
classify our introduced \gls{ibpt} Industry~4.0 scenario.

\subsection{Vita perspective}
\label{subsec:vita-perspective}

\begin{figure}[tb]
    \centerline{\includegraphics[width=1.0\linewidth, page=7]{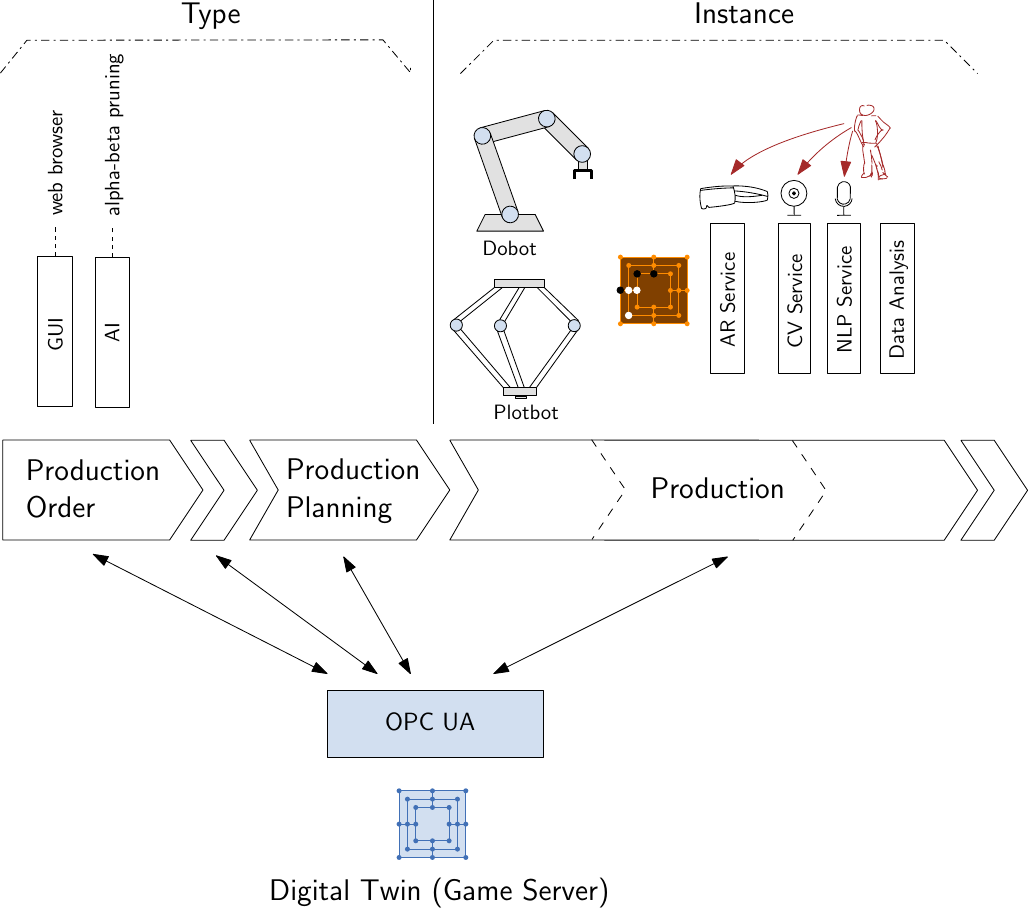}}
    \caption{\emph{Vita} view of Nine Men's Morris production order process.}
    \label{fig:vita}
\end{figure}

The \emph{Vita} perspective describes how every object in an industrial value
chain goes through different phases, which are divided into the two major
segments \emph{Type} and \emph{Instance}.
The \emph{Type} segment describes the development phase of the object, while
the \emph{Instance} segment describes how the object is produced, used and
disposed.
This principle is shown in Fig.~\ref{fig:vita}, along with the life cycle of
a production order of our Industry~4.0 scenario of playing Nine Men's Morris.

In this scenario, the object we want to produce is a certain game state.
During the \emph{Type} phase, a production order can be placed in the form of game moves.
The \emph{Game Server} validates the order and
distributes the move to its connected production units, which are the
geographically distributed robot cells.

As shown in Fig.~\ref{fig:vita}, information is exchanged throughout the
\emph{Type} and \emph{Instance} phase between actors along the value chain and
the \gls{ibpt} entity. In this way, every entity of involved \gls{ot} and
\gls{it} systems is linked to one or several phases of the business process via
the \gls{ibpt}. During production, changes made to the game state (and therefore
the product) are recognized by \gls{nlp}, \gls{ai} or \gls{ar}, triggering yet
another production order. In the sense of \gls{uml} sequence diagrams, in this
way, the lifelines of production orders can exist simultaneously and information  
can be exchanged between different production orders and phases.
Dividing the \emph{Production} phase into sub-phases (dashed lines), like, e.g.
\enquote{pick up token}, \enquote{move token} and \enquote{place token}, allows
for the \gls{ibpt} entity to collect specific data of each phase in order to
store this data for later use. 

\subsection{Hierarchical perspective}
\label{subsec:hierarchical-perspective}

\begin{figure}[tb]
    \centerline{\includegraphics[width=1.0\linewidth, page=5]{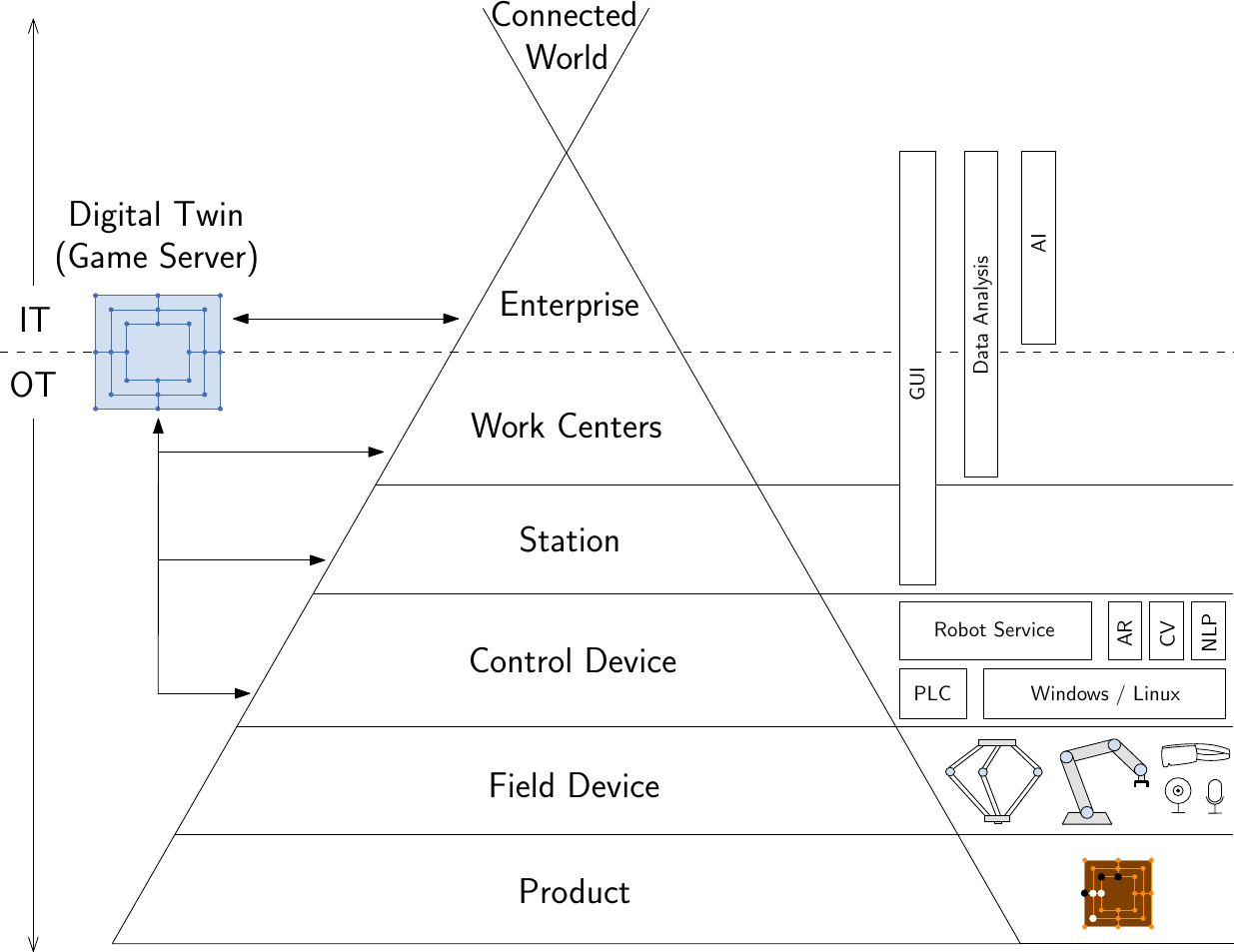}}
    \caption{Nine Men's Morris \gls{ibpt} \emph{Hierarchical} view.}
    \label{fig:hierarchical}
\end{figure}

The \emph{Hierarchical} perspective depicted in Fig.~\ref{fig:hierarchical}
shows vertical integration within a production facility. It closely resembles
the hierarchical level approach of the ISA-95 pyramid, extending it by layers
\emph{Connected World} and \emph{Product}. Our \gls{ibpt} entity is linking the
worlds of \gls{it} and \gls{ot} by interacting with \gls{ot} \emph{Work
Centers}, \emph{Station} and \emph{Control Device} levels as well as \gls{it}
services.

The game state product manifests in the form of token positions on the physical
game board located on the \emph{Product} hierarchy level.
Production units located on \emph{Field Device} level establish these token
positions.
They are controlled via devices on the \emph{Control Hardware} level, which
receive game moves from the \gls{ibpt} entity.
A connection to other production sites is established via \gls{opcua} over the
Internet.

\subsection{Architectural perspective}
\label{subsec:architectural-perspective}

\begin{figure}[tb]
    \centerline{\includegraphics[width=1.0\linewidth, page=8]{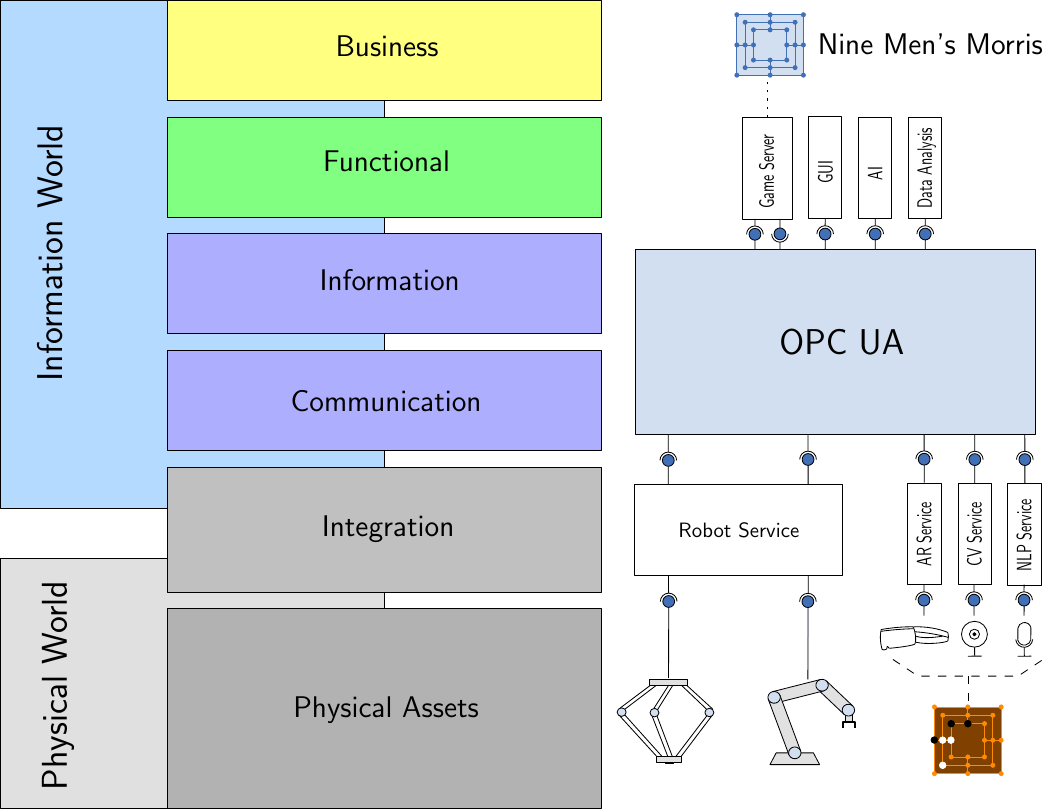}}
    \caption{Nine Men's Morris \gls{ibpt} \emph{Architectural} view.}
    \label{fig:architectural}
\end{figure}

Physical assets are represented on the lowest level of the \emph{Architectural}
perspective shown in Fig.~\ref{fig:architectural}.
These assets of the physical world are connected to the information world via
devices and programs running at the \emph{Integration} architectural layer.
Physical assets include robots and \gls{hmi} input devices, as well as the physical game
boards with respective game states located in each robot cell.
\gls{ar} glasses, a \gls{cv} camera or \gls{nlp} microphone detect the state of
the game board and, via services located at the \emph{Integration} layer level,
propagate this state into the information world.

\gls{opcua} builds the common middleware for information modeling and
communication, linking services running at the \emph{Functional} layer with the
business process of playing the game of Nine Men's Morris as well as services
running at the \emph{Integration} layer.

\section{Conclusion and Outlook}
\label{sec:conclusion-and-outlook}

We have introduced an \gls{ibpt} entity based on a business-centric \gls{soa}
architecture. The proposed entity acts as an intermediary, decoupling the worlds
of \gls{it} and \gls{ot}, thus allowing for an integration of \gls{it} and
\gls{ot} components of different manufacturers and platforms. 
Methods of one world, in this way, are not directly applied to the other, but,
instead, to a representation, that is in bidirectional exchange with the other
world. 
Using this \gls{ibpt} approach, we have demonstrated the four essential
Industry~4.0 design principles described by Hermann et al.\@ in
\cite{Hermann2016} based on the gamified Industry~4.0 scenario of playing the
game of Nine Men's Morris.
This game-based scenario serves well for agent based \gls{ai}-research and
education.

In future work, the \gls{ibpt} entity could be realized as Industry~4.0
component compliant to the definition provided in~\cite{Heidel2019}. This
requires the extension of our data model to meet the requirements of
\glspl{aas}. Specifications are provided by the \enquote{Plattform
Industrie~4.0} association in~\cite{platformindustry402022_1}
and~\cite{platformindustry402022_2} and define the base \gls{aas} model as well
as functional \glspl{api}. Specific data acquired by the \gls{ibpt} throughout
each phase of a production order can be stored in submodels and made available
to the Industry~4.0 system for data analysis. In this way, as an example, by
storing data on positioning accuracy using the \gls{cv} component for each
respective production phase, one can later determine whether positioning
deviations during the \enquote{pick up token} phase propagate to the
\enquote{place token} phase via standardized access to \gls{aas} submodels.

In~\cite{Pauker2016}, Pauker et al.\@ demonstrate a method to directly transform
\gls{uml} diagrams into \gls{opcua} models. This approach could be extended in
order for the generated asset and \gls{aas} model to meet the specification
of~\cite{platformindustry402022_1} and~\cite{platformindustry402022_2} and
therefore allowing for an automated generation of Industry~4.0 compliant
\gls{ibpt} information models. This could be accompanied by business process
modeling standards like the \gls{bpmn}.

\section*{Acknowledgements}

We want to thank Arndt Lüder for his valuable input on Industry~4.0 compliant information modeling.

\bibliographystyle{IEEEtran}
\bibliography{references}

\end{document}